# From Bench-to-Bedside: A Review of Clinical Trials in Drug Discovery and Development

## Graphical abstract

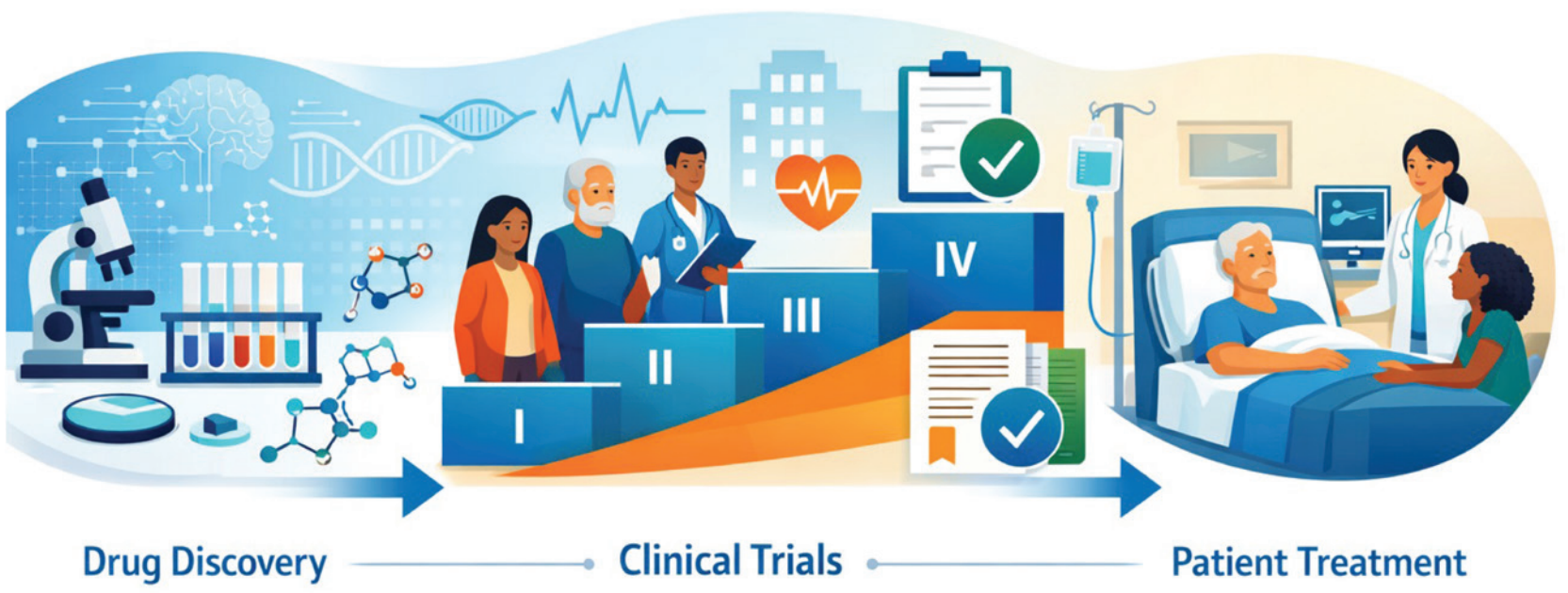

## Highlights

- Clinical trials progress through four sequential phases (I–IV), each addressing distinct safety and efficacy questions in drug development.
- Adaptive trial designs and Bayesian approaches enable real-time modifications, improving efficiency and reducing development timelines.
- Artificial intelligence and machine learning are transforming patient recruitment, trial monitoring, and data analysis processes.
- Emerging therapies, including gene therapy and immunotherapy, require innovative trial designs with biomarker-driven enrollment strategies.
- International regulatory harmonization and decentralized trial capabilities are essential for global clinical development programs.

## Authors

Tianyang Wang, Ming Liu, Benji Peng, Xinyuan Song, Charles Zhang, Xintian Sun, Qian Niu, Junyu Liu, Silin Chen, Keyu Chen, Ming Li, Pohsun Feng, Ziqian Bi, Yunze Wang, Yichao Zhang, Cheng Fei, Lawrence K. Q. Yan, Ziyuan Qin, Riyang Bao and Zekun Jiang

## Correspondence

zekun_jiang@163.com
(Z. Jiang)

## Brief statement

This comprehensive review synthesizes current knowledge on clinical trial methodology, technological innovations, and regulatory frameworks, providing actionable insights for researchers, sponsors, and policymakers to accelerate drug development, while maintaining scientific rigor and patient safety.

# From Bench-to-Bedside: A Review of Clinical Trials in Drug Discovery and Development

Tianyang Wang[1], Ming Liu[2], Benji Peng[3,4], Xinyuan Song[5], Charles Zhang[4], Xintian Sun[6], Qian Niu[7], Junyu Liu[7], Silin Chen[8], Keyu Chen[4], Ming Li[4], Pohsun Feng[9], Ziqian Bi[10], Yunze Wang[11], Yichao Zhang[12], Cheng Fei[13], Lawrence K. Q. Yan[14], Ziyuan Qin[5], Riyang Bao[5] and Zekun Jiang[15,*]

[1]University of Liverpool, Liverpool, UK

[2]Purdue University, West Lafayette, USA

[3]Georgia Institute of Technology, Atlanta, USA

[4]AppCubic, Miami, USA

[5]Emory University, Atlanta, USA

[6]Simon Fraser University, Burnaby, Canada

[7]Kyoto University, Kyoto, Japan

[8]Zhejiang University, Hangzhou, China

[9]National Taiwan Normal University, Taipei City, Taiwan, China

[10]Indiana University, Bloomington, USA

[11]University of Edinburgh, Edinburgh, UK

[12]The University of Texas at Dallas, Richardson, USA

[13]University of Wisconsin Madison, Madison, USA

[14]The Hong Kong University of Science and Technology, Hong Kong, China

[15]West China Biomedical Big Data Center, West China Hospital, Sichuan University, Chengdu, Sichuan, China

*Correspondence to: Zekun Jiang, West China Biomedical Big Data Center, West China Hospital, Sichuan University, No. 37 Guoxue Alley, Wuhou District, Chengdu 610000, China. E-mail: zekun_jiang@163.com

## Abstract

Clinical trials bridge basic research and clinical application, serving as essential steps in drug development. This review examines clinical trial phases (Phase I [safety assessment], Phase II [efficacy evaluation], Phase III [large-scale validation], and Phase IV [post-marketing surveillance]), highlighting the distinct characteristics and interconnections. Major challenges are identified, including ethical compliance, participant recruitment, and ensuring diversity and representativeness in trial populations, while proposing evidence-based mitigation strategies. To address these challenges, innovative technologies, such as artificial intelligence, big data analytics, and digital health tools, are transforming trial design and implementation, enhancing efficiency and data quality. Looking forward, the review explores how emerging therapies, including gene therapy and immunotherapy, are reshaping trial design requirements and emphasizes the growing importance of regulatory harmonization and global collaboration. Clinical trials remain central to advancing innovative drug development and improving patient outcomes.

## Introduction

Clinical trials serve as the critical bridge connecting laboratory discoveries to clinical practice, representing the cornerstone of evidence-based drug development [1, 2]. These rigorously designed studies generate the scientific evidence required by regulatory authorities to evaluate whether new therapeutics should be approved for patient use [3]. Without clinical trials promising laboratory findings would remain untranslated and unable to benefit the millions of patients worldwide who depend on innovative treatments for major diseases [4].

The clinical trial framework comprises four sequential phases, each addressing specific questions in drug development [5, 6]. Phase I trials, typically enrolling 20–100 healthy volunteers, establish safety profiles and determine appropriate dosing ranges. Phase II trials expand to 100–300 patients with the target condition to evaluate preliminary efficacy and optimize dosing regimens. Phase III trials involve 300–3000 participants across multiple centers to confirm therapeutic benefits and monitor adverse effects in diverse populations. Finally, Phase IV post-marketing surveillance continues after regulatory approval, detecting rare adverse events and assessing long-term outcomes in real-world settings [7].

Despite an essential role, clinical trials face substantial challenges that contribute to high failure rates and escalating costs [8]. Ethical considerations demand rigorous informed consent processes and institutional oversight to protect participant welfare. Patient recruitment remains a persistent bottleneck with many trials failing to meet enrollment targets due to strict eligibility criteria, geographic barriers, and patient concerns about experimental treatments [9, 10]. The globalization of clinical research has introduced additional complexity, requiring navigation of diverse regulatory frameworks, cultural contexts, and healthcare systems [11].

Recent years have witnessed transformative technological advances reshaping clinical trial conduct [12, 13]. Artificial intelligence and machine learning now support patient identification, trial design optimization, and real-time safety monitoring. Digital health technologies, including wearable devices and mobile applications, enable remote data collection and decentralized trial models that reduce participant burden [14]. Real-world evidence derived from electronic health records and claims databases increasingly supplements traditional trial data, providing insights into treatment effectiveness across diverse patient populations [15].

The emergence of novel therapeutic modalities, including gene therapies, cell-based treatments, and immunotherapies, poses unique challenges for conventional trial designs [16, 17]. These innovative approaches often target small patient populations with rare diseases, require personalized manufacturing processes, and necessitate extended follow-up periods to assess durability of response. Precision medicine principles increasingly guide trial design with biomarker-driven enrollment strategies enabling more efficient identification of responsive patient subgroups [18].

Regulatory frameworks continue to evolve in response to scientific advances and globalization pressures [19, 20]. Expedited approval pathways, including breakthrough therapy designations and accelerated approval mechanisms, aim to bring promising treatments to patients faster, while maintaining safety standards. However, international regulatory harmonization remains incomplete, creating challenges for multinational trials and global drug development programs. The future of clinical trials will require balancing innovation with rigor, leveraging new technologies, while ensuring robust evidence generation.

This review provides a comprehensive examination of clinical trials in drug discovery and development with emphasis on three integrated themes: (1) methodologic foundations of trial design across all phases; (2) technological innovations, including artificial intelligence, real-world evidence, and digital health tools; and (3) regulatory considerations and strategies for addressing persistent challenges. Unlike reviews focusing on individual aspects, this work synthesizes methodologic, technological, and regulatory perspectives to provide actionable insights for researchers, sponsors, regulators, and policymakers. The opportunities and limitations of emerging approaches are critically evaluated, identifying gaps between technological promise and practical implementation in the evolving clinical trial landscape.

## Literature search strategy

This review was conducted following a structured literature search methodology. PubMed, Web of Science, Scopus, and Cochrane Library databases were searched for articles published between January 2015 and December 2024. Search terms included combinations of “clinical trials,” “drug development,” “trial design,” “regulatory approval,” “artificial intelligence,” “adaptive design,” and “personalized medicine.” Peer-reviewed original research, systematic reviews, meta-analyses, and regulatory guidance documents published in English were included. Seminal historical works and landmark trial reports were included regardless of publication date to provide foundational context. Reference lists of the included articles were manually screened to identify additional relevant publications. In total, more than 150 articles were reviewed with 95 meeting inclusion criteria for citation in this manuscript.

## Early stages of drug discovery

### Target identification and validation

The first critical step in drug discovery is the identification and validation of potential drug targets [21]. Drug targets are typically proteins, enzymes, or receptors that have key roles in disease processes. The process begins with understanding the biological mechanisms underlying a particular disease, often through genomic, proteomic, and bioinformatics techniques [22]. Once a potential target is identified, the target must be validated to ensure that modulating activity will have a therapeutic effect. Target validation can be achieved through various methods, including genetic manipulation (e.g., gene knockouts or knockdowns), pharmacologic inhibition, or the use of animal models that mimic the disease condition [23]. Only after thorough validation can a target move forward to the next stages of drug discovery [24].

### Hit discovery and screening

After target identification and validation, the next step is to discover small molecules, biologics, or other entities that can interact with the target effectively [25]. This process is known as hit discovery, which involves screening large chemical libraries or using computational methods like virtual screening. High-throughput screening (HTS) is a widely used method that tests thousands of compounds against the target to identify “hits” (compounds that show desired activity). In addition to chemical libraries, natural products and biologics, including monoclonal antibodies or peptides, may also be investigated for potential as therapeutic agents [26, 27]. The identified hits are then subjected to further testing for potency, selectivity, and the potential for off-target effects.

### Lead optimization

Following identification, hits undergo a process called lead optimization. During this stage, the goal is to refine the chemical structure of the hit compounds to improve the potency, selectivity, pharmacokinetics, and toxicity profiles [28]. Medicinal chemistry has a crucial role in optimizing leads to enhance drug-like properties. Techniques, such as structure-activity relationship (SAR) studies, molecular modeling, and iterative chemical synthesis, are used to modify the compounds [29]. This stage may involve the synthesis of hundreds or thousands of analogs, followed by biological testing to identify the best candidates for preclinical development. Lead optimization is crucial to developing drug candidates with the desired characteristics for clinical trials; failure to optimize effectively can lead to failure of the drug at later stages [30].

## Preclinical development

Preclinical development involves evaluating the drug candidate in non-human models to assess the safety, pharmacokinetics, and efficacy before moving to human trials [31]. This phase includes a variety of *in vitro* and *in vivo* studies to determine the absorption, distribution, metabolism, excretion, and toxicity (ADMET) properties of the compound. Animal studies are also used to establish the preliminary therapeutic window and dosing regimen. The results of preclinical studies are critical in informing decisions regarding the appropriate dose range, treatment schedule, and potential side effects in humans [32, 33]. Regulatory agencies, such as the US Food and Drug Administration (FDA) and European Medicines Agency (EMA), require extensive preclinical data to approve the initiation of clinical trials, ensuring that the drug candidate is safe enough for human testing [34, 35].

The transition from preclinical development-to-clinical trials represents a critical decision point in drug development. An Investigational New Drug (IND) application must be submitted to regulatory authorities that contains comprehensive preclinical data on pharmacology, toxicology, and manufacturing. The IND serves as the bridge connecting laboratory findings to human testing, the approval of which marks the beginning of the clinical evaluation journey described in the following section.

# Clinical trial phases

Following successful preclinical development and IND approval, drug candidates enter the clinical trial phase, which is a systematic evaluation process designed to establish safety and efficacy in humans. The clinical trial framework consists of four sequential phases, each with distinct objectives, patient populations, and regulatory requirements. This progression from first-in-human studies to post-marketing surveillance represents the "problem-solution" approach central to drug development, as follows: Phase I answers the safety question; Phase II answers the efficacy question; Phase III answers the generalizability question; and Phase IV answers the long-term safety question.

## Phase I: safety and dose exploration

Phase I clinical trials are the first step in testing a new drug in humans. The primary focus of Phase I is to assess the safety, tolerability, and pharmacokinetics of the investigational drug. Typically, these trials are conducted with a small group of healthy volunteers (20–100 individuals). The main objective is to determine the appropriate dosage range for future studies [36, 37]. This phase also aims to identify any potential side effects, the ADME properties of the drug, as well as the most effective route of administration [38].

Dose escalation studies are commonly used in Phase I trials, in which the drug dose is gradually increased to determine the maximum tolerated dose (MTD) [39–41]. The findings from this phase are crucial in providing foundational data on the safety profile of the drug, which can then be used to design the subsequent phases.

## Phase II: preliminary efficacy and dose optimization

Phase II trials primarily focus on evaluating the efficacy of the drug in patients who have the target disease or condition. Phase II trials aim to determine whether the drug shows signs of therapeutic benefit, typically involving a larger cohort (100–300 patients). These trials also continue to monitor safety and adverse effects, particularly the effects that may occur at therapeutic doses [42–44].

Phase II trials are often divided into two subphases: the Phase IIa subphase focuses on pharmacodynamic effects and early indications of therapeutic responses, and helps establish the optimal dose range for treatment; and the Phase IIb subphase further investigates the efficacy of the drug in a larger patient population and helps refine the dose-response relationship; data from Phase IIb trials are essential for deciding whether the drug progresses to Phase III [45].

The results of Phase II trials provide vital information on the potential of the drug to treat the disease and safety at the target doses.

## Phase III: large-scale validation and market access

Phase III clinical trials are the most critical for regulatory approval and the eventual market introduction of a drug. The drug is tested in a large patient population (typically 300–3000 participants) in Phase III to confirm efficacy and further assess the safety profile [46–48]. These trials are often multicenter studies and involve diverse patient populations to ensure the results are generalizable across different demographics [20, 49].

Phase III trials are designed to provide definitive evidence of drug benefits compared to existing treatments or a placebo. Regulatory agencies, such as the US FDA or the EMA, require comprehensive data from Phase III trials to assess whether the drug should be approved for public use [50, 51]. The success of Phase III trials is the final step before a New Drug Application (NDA) or Marketing Authorization Application (MAA) is submitted [52, 53].

The safety profile continues to be assessed during this phase, particularly regarding rare or long-term side effects, which may not have been identified in earlier phases [49, 54]. The findings from Phase III trials have a key role in the market access and widespread use of the drug.

### Critical perspective: learning from trial controversies

The history of Phase III trials includes both triumphs and cautionary tales that have shaped current regulatory practices. The withdrawal of rofecoxib (Vioxx) in 2004 following cardiovascular safety signals that emerged only after widespread market use highlighted fundamental limitations in detecting rare adverse events even in large Phase III populations [55]. The APPROVe trial revealed that patients taking Vioxx had twice the risk of a myocardial infarction and stroke compared to placebo, yet these signals were not apparent in the original Phase III trials that enrolled approximately 5000 patients. This case prompted the FDA Risk Evaluation and Mitigation Strategies (REMS) program and increased emphasis on cardiovascular safety trials for COX-2 inhibitors and similar drug classes.

Conversely, the accelerated development of trastuzumab (Herceptin) for HER2-positive breast cancer exemplifies how biomarker-driven trial design can identify responsive patient subpopulations and achieve remarkable efficacy [56]. By restricting enrollment to patients with HER2 overexpression (approximately 25% of breast cancer patients), the pivotal Phase III trial demonstrated a 20% improvement in survival, an effect size that would have been diluted to insignificance in an unselected population. This success established the paradigm for precision medicine approaches in oncology and demonstrated that smaller, biomarker-selected trials can achieve regulatory approval with compelling efficacy data.

## Phase IV: post-marketing surveillance and long-term effects

Phase IV trials, also known as post-marketing surveillance, take place after a drug has received regulatory approval and is available to the general public [57, 58]. These trials are designed to monitor the long-term safety and efficacy of the drug in the broader population. While Phase I–III trials provide essential data on drug safety and effectiveness, Phase IV trials allow researchers to observe drug performance over a longer period and in more diverse patient populations [59, 60].

Post-marketing surveillance can identify rare or unexpected adverse effects that might not have been evident in earlier trials due to smaller sample sizes. In addition, Phase IV studies may explore new therapeutic indications, drug interactions, or effectiveness when combined with other treatments [61]. These studies are also critical for continuing to ensure that the benefits of the drug outweigh any potential risks, especially as the drug is used by a larger and more varied patient population.

## Interconnectivity and translational aspects of the phases

The four phases of clinical trials are interrelated and build upon each other phase to ensure that a new drug is both effective and safe for use in the general population [62, 63]. Phase I provides the foundational safety data that guide the design of Phase II trials, in which the primary focus shifts toward assessing therapeutic efficacy and refining dosing. Phase II results inform Phase III, in which large-scale validation occurs to confirm the overall benefits and safety of the drug in a broader population [5, 64, 65]. Finally, Phase IV studies provide long-term monitoring to assess real-world effectiveness and safety.

## Adaptive designs and Bayesian approaches

Modern clinical trials increasingly use adaptive designs that allow pre-specified modifications based on accumulating data [66]. Adaptive trials can modify randomization ratios, drop ineffective treatment arms, or re-estimate sample sizes at interim analyses unlike traditional fixed designs, while maintaining statistical validity. These designs address the inefficiencies of conventional trials, in which approximately 50% of Phase II and 40% of Phase III trials fail despite substantial investment.

**Key Adaptive Design Types:**

- **Adaptive Randomization:** Response-adaptive randomization assigns more patients to better-performing treatment arms, improving both ethical considerations and statistical efficiency. The I-SPY 2 trial in breast cancer demonstrated that this approach can reduce the number of patients exposed to inferior treatments, while accelerating identification of effective therapies.
- **Seamless Phase II/III Designs:** These designs combine dose-finding and confirmatory phases into a single trial, reducing development timelines by 12–18 months. Data from the Phase II portion contribute to the final analysis, improving statistical power without increasing overall sample size [45].
- **Platform Trials:** Master protocols, such as I-SPY 2 and RECOVERY, allow continuous evaluation of multiple treatments against a common control with treatments entering and exiting based on pre-defined success or futility criteria [67]. The rapid identification of dexamethasone efficacy during COVID-19 in the RECOVERY trial demonstrated the potential of the platform trial for accelerating therapeutic discovery during public health emergencies.

**Bayesian Methods in Clinical Trials:** Bayesian approaches enable incorporation of prior knowledge and continuous updating of treatment effect estimates [62]. Bayesian designs integrate historical information and expert knowledge through prior distributions, unlike frequentist methods that rely solely on trial data. The US FDA has increasingly accepted Bayesian designs, particularly for medical devices and rare diseases in which traditional frequentist approaches requiring large sample sizes are impractical. Key advantages include more intuitive probability statements about treatment effects, efficient use of accumulating data, and natural handling of interim analyses without multiplicity penalties.

**Digital Biomarkers:** Wearable devices and smartphone sensors generate continuous physiologic data that complement traditional clinical endpoints [14]. Digital biomarkers, such as gait speed measured via accelerometers, sleep patterns from wearable devices, and heart rate variability provide objective, are real-world measures of treatment effects. The Parkinson's Progression Markers Initiative (PPMI) demonstrated that smartphone-based assessments can detect motor function changes with greater sensitivity than quarterly clinic visits. However, validation standards and regulatory acceptance remain evolving areas with US FDA guidance on digital health technologies continuing to develop.

## Challenges and response strategies in clinical trials

Clinical trials continue to face persistent challenges that compromise efficiency, validity, and generalizability, despite significant advances in trial methodology and technology. These challenges are interconnected; ethical concerns affect recruitment willingness, recruitment difficulties exacerbate diversity gaps, and inadequate diversity undermines data quality and regulatory acceptance. Understanding these interdependencies is essential for developing effective mitigation strategies.

The magnitude of these challenges is substantial. Approximately 85% of clinical trials fail to meet recruitment targets within planned timeframes, while dropout rates average 30% across all trial phases [68]. These failures translate into significant financial consequences that extend trial timelines by just 1 d costs sponsors an estimated $37,000 in direct expenses with total delays potentially adding millions of dollars to development costs [69]. More critically, prolonged trials delay patient access to potentially life-saving treatments and reduce the statistical power needed to detect meaningful clinical effects.

**Figure 1** presents a hierarchical framework organizing these challenges into four interconnected categories: (1) ethical and regulatory compliance, which establishes the foundational requirements for trial conduct; (2) patient recruitment and retention, which determines whether sufficient participants can be enrolled and maintained; (3) diversity and population representativeness, which affects the generalizability of findings; and (4) data quality and operational compliance, which ensures the reliability of results. The following subsections examine each category, analyzing specific obstacles and presenting evidence-based strategies to address the categories.

## Ethical and regulatory compliance

Ethical and regulatory compliance forms the foundation of credible clinical research. Historical failures in research ethics, including the Tuskegee syphilis study and thalidomide-related birth defects, have shaped modern regulatory frameworks and underscored the necessity of rigorous participant protections [70]. Today, the core ethical challenge remains balancing scientific advancement with participant welfare, requiring researchers to navigate complex requirements, while maintaining trial feasibility.

**Informed Consent Challenges:** Informed consent represents the cornerstone of ethical research, requiring participants to voluntarily agree to enrollment after understanding the purpose, risks, benefits, and alternatives of the trial [71]. However, implementing meaningful informed consent faces practical obstacles. Consent documents have grown increasingly complex, often exceeding 20 pages with technical language that participants struggle to comprehend. Studies indicate that many participants sign consent forms without fully understanding key elements, particularly regarding randomization and placebo use. Cultural and linguistic differences further complicate consent in multinational trials, in which concepts, like "randomization," may lack direct translations or conflict with local healthcare expectations [72].

**Regulatory Complexity:** Clinical trials must comply with regulations from multiple authorities, including the US FDA, EMA, and national agencies worldwide [73, 74]. While these agencies share common objectives, the specific requirements often differ in areas, such as pediatric study mandates, ethnic bridging studies, and data submission formats. Multinational sponsors must navigate these differences, sometimes conducting additional studies or modifying protocols to satisfy regional requirements [75]. The regulatory burden is particularly acute for academic investigators and smaller sponsors that lack dedicated regulatory affairs departments.

**Privacy and Data Protection:** The expansion of digital data collection and international data sharing has intensified privacy concerns. Regulations, such as General Data Protection Regulation (GDPR) in Europe and the Health Insurance Portability and Accountability Act (HIPAA) in the United States impose strict requirements on data handling, while emerging frameworks in other regions add complexity for global trials. Balancing data utility for research with participant privacy requires sophisticated governance approaches that many organizations are still developing.

**Response Strategies:**

- **Simplify and enhance informed consent:** Plain-language consent documents supplemented with visual aids and multimedia explanations are developed. Staged consent processes that revisit key information are considered at multiple timepoints [76].
- **Adopt harmonized regulatory approaches:** International Council for Harmonisation (ICH) guidelines and mutual recognition agreements are leveraged to streamline multinational submissions. Regulatory agencies are engaged early through pre-submission meetings to clarify requirements.
- **Implement robust data governance:** Comprehensive data protection frameworks addressing collection, storage, transfer, and retention are established. Privacy-enhancing technologies, such as anonymization and federated analysis, are utilized as appropriate.
- **Build regulatory expertise:** An investment in training for research staff is made and partnerships with regulatory consultants for complex submissions are considered.

## Patient recruitment and retention

Recruitment and retention represent the most common operational barriers to successful trial completion. Industry data indicate that 85% of trials fail to recruit sufficient participants within planned timeframes, while approximately 30% of enrolled participants discontinue before study completion [68, 77]. These challenges vary substantially by therapeutic area; oncology trials typically achieve higher enrollment rates due to limited treatment alternatives, while trials for chronic conditions with existing therapies face greater recruitment difficulties [78].

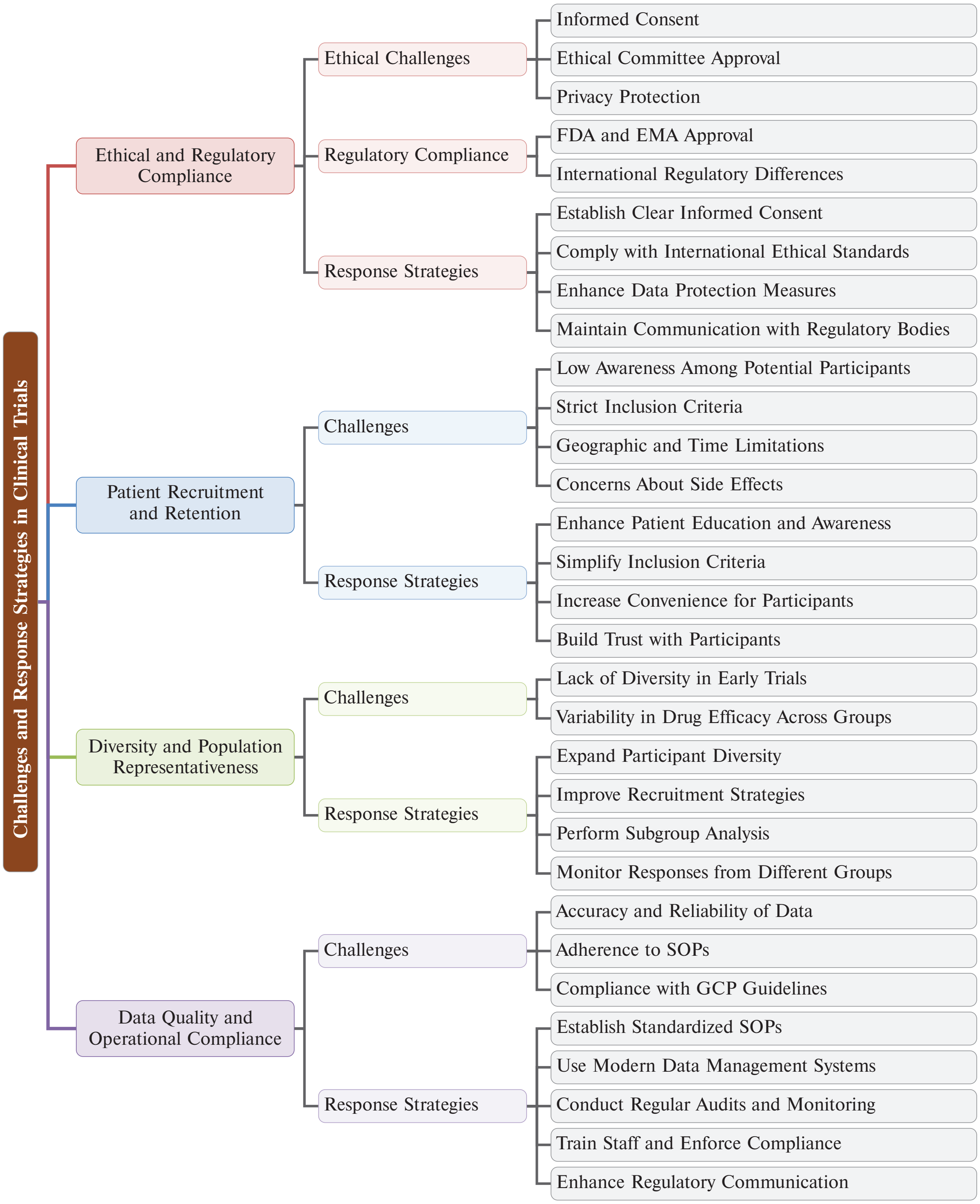

**Figure 1** Hierarchical overview of clinical trial challenges and mitigation tactics.

**Recruitment Barriers**: Multiple factors impede patient enrollment. Strict eligibility criteria exclude substantial proportions of screened patients; screen failure rates exceed 35% across all trial types and approach 50% in some therapeutic areas. Many patients remain unaware of relevant trials or hold misconceptions about participation, including fears of receiving placebos when effective treatments exist. Geographic barriers persist despite digital advances because many trials concentrate at academic medical centers accessible primarily to urban populations. Financial concerns, including lost wages and travel costs, disproportionately affect lower-income populations [79].

**Retention Challenges**: Maintaining participant engagement throughout trial duration poses distinct challenges. Common dropout triggers include adverse events (real or perceived), treatment burden from frequent visits or complex protocols, loss of perceived benefit, and life circumstances, such as relocation or competing health priorities. The COVID-19 pandemic highlighted these vulnerabilities with survey data showing that 44% of participants found travel to study sites "somewhat" or "very burdensome" in 2021, a 15% increase from 2019 [80]. Protocol complexity correlates strongly with dropout; trials requiring more frequent visits, longer duration, or invasive procedures experience higher attrition.

**Disease-specific Considerations**: Recruitment and retention patterns differ markedly across therapeutic areas. Rare disease trials face inherently limited patient pools but often benefit from highly motivated participants and engaged patient advocacy organizations. Pediatric trials require parental consent and child assent, which adds complexity, while facing restrictions on acceptable risk levels. Psychiatric trials contend with stigma-related barriers and cognitive factors affecting consent capacity. Oncology trials increasingly compete for patients as treatment options multiply, while cardiovascular prevention trials must convince asymptomatic individuals to participate in long-duration studies.

**Response Strategies:**

- **Optimize eligibility criteria:** "Fit-for-purpose" approaches that balance scientific rigor with real-world practicality are applied. Broader criteria improve recruitment, while potentially enhancing the generalizability of the findings.
- **Implement decentralized trial elements:** Telemedicine visits, home nursing services, and direct-to-patient drug delivery are utilized to reduce participant burden. Wearable devices and mobile applications enable remote data collection, while maintaining data quality.
- **Enhance participant engagement:** Comprehensive communication strategies addressing participant questions and concerns throughout the trial are developed. Regular updates on study progress help participants feel valued and connected to the research mission.
- **Address financial barriers:** Adequate compensation for time and travel are provided, and insurance coverage policies that do not penalize trial participation are advocated.
- **Leverage technology for recruitment:** Electronic health record screening, social media outreach, and patient registry partnerships are utilized to identify and engage potential participants more efficiently.

## Diversity and population representativeness

The lack of diversity in clinical trials represents both a scientific limitation and an ethical concern. Historically, clinical trials predominantly enrolled white males, leaving significant knowledge gaps about drug effects in women, elderly patients, and racial/ethnic minorities [81]. This underrepresentation has tangible consequences; drugs may receive approval based on data that inadequately predict responses in substantial portions of the intended patient population [82].

Biological and Clinical Implications: Pharmacokinetic and pharmacodynamic differences across demographic groups are well-documented. Women generally have higher body fat percentages than men, which affects drug distribution and slows gastric emptying. As a result, absorption is affected and the expression of drug-metabolizing enzymes differs. Genetic polymorphisms affecting drug metabolism vary significantly by ancestry. For example, CYP2D6 poor metabolizer phenotypes occur in approximately 7% of Caucasians but only 1% of East Asians, while CYP2C19 poor metabolizers are more common in Asian populations. These differences can substantially affect drug efficacy and toxicity, as demonstrated by the differential responses to clopidogrel across ethnic groups and the higher incidence of ACE inhibitor-induced angioedema in Black patients.

Regulatory Evolution: Recognition of these gaps has prompted regulatory action. The FDA 2024 draft guidance on Diversity Action Plans, mandated by the Food and Drug Omnibus Reform Act of 2022 (FDORA), requires sponsors to submit plans specifying enrollment goals disaggregated by race, ethnicity, gender, and age for certain clinical studies [83]. Sponsors must provide rationale for the enrollment targets and describe strategies for achieving the enrollment targets. The guidance emphasizes that meaningful diversity requires proactive planning rather than retrospective justification of homogeneous enrollment.

Structural Barriers to Diversity: Achieving representative enrollment requires addressing systemic barriers. Clinical trial sites concentrate in academic medical centers serving predominantly White, insured populations. Historical exploitation of minority communities in research, as exemplified by the Tuskegee study, has created lasting distrust that affects willingness to participate. Language barriers, inflexible scheduling around work hours, and lack of transportation disproportionately affect underrepresented populations. In addition, eligibility criteria often exclude patients with co-morbidities that are more prevalent in minority populations, such as hypertension or diabetes.

**Response Strategies:**

- **Develop prospective diversity plans:** Specific enrollment targets are established for underrepresented populations at the trial design stage. Enrollment demographics are monitored throughout the study and corrective actions are implemented when targets are not being met.
- **Expand site networks:** Community health centers, safety-net hospitals, and practices serving diverse populations are included, partnering with minority-serving institutions and engaging with community advisory boards.
- **Address practical barriers:** Flexible scheduling is offered, transportation assistance is provided, materials are available in relevant languages, and staff are trained in cultural competency [84, 85].
- **Build community trust:** Patient advocacy organizations representing diverse communities are engaged. Trial results are made available to participating communities. Diverse representation is assured among research staff and investigators [86].
- **Conduct meaningful subgroup analyses:** Studies are adequately powered to detect clinically meaningful differences across demographic subgroups. Transparent subgroup results are reported, including limitations when subgroups are underpowered.

## Data quality and operational compliance

Data quality and operational compliance determine whether trial results can support regulatory decisions and withstand scientific scrutiny. Poor data quality can invalidate otherwise well-designed studies, waste resources, and delay patient access to effective treatments [87]. Regulatory agencies have increasingly emphasized quality management approaches that focus on factors critical to participant safety and data reliability.

Sources of Data Quality Issues: Data quality problems arise from multiple sources throughout the trial lifecycle. Protocol deviations, including missed visits, incorrect dosing, and eligibility violations, occur in most trials and can compromise interpretability if not properly managed. Data entry errors in case report forms remain common despite electronic capture systems. Source data verification during monitoring visits reveals discrepancies between source documents and reported data in a significant proportion of records. More fundamentally, unclear protocol instructions and inadequate site training contribute to systematic data collection errors that affect entire datasets [88].

Compliance Challenges in Modern Trials: The Good Clinical Practice (GCP) framework provides internationally harmonized standards for trial conduct but implementation faces ongoing challenges [89]. Decentralized trial elements, including telemedicine visits, wearable devices, and direct-to-patient drug delivery, introduce new compliance considerations not fully addressed by traditional GCP frameworks developed for site-based trials. International trials must navigate varying interpretations of GCP requirements across regulatory jurisdictions. In addition, increasing protocol complexity has outpaced site capabilities with > 40% of trials requiring protocol amendments before the first participant visit.

Emerging Quality Approaches: Regulatory agencies have shifted from exhaustive verification toward risk-based quality management. This approach prioritizes monitoring and quality activities on factors most likely to affect participant safety and data reliability for critical decisions. Central statistical monitoring can identify data anomalies suggesting fraud, systematic errors, or site performance issues more efficiently than traditional 100% source data verification. However, implementing risk-based approaches requires sophisticated data analytics capabilities that many organizations are still developing.

**Response Strategies:**

- **Implement risk-based quality management:** Critical data and processes are identified at the trial design stage. Quality activities focus on protecting participant safety and ensuring reliability of data supporting primary endpoints [90].
- **Leverage technology for quality:** Electronic data capture is deployed with built-in validation checks and real-time data review. Central monitoring platforms that apply statistical algorithms are utilized to detect data anomalies across sites [91].
- **Simplify protocols:** Unnecessary complexity that increases error rates and site burden is reduced. Data collection is limited to information essential for answering the research question and ensuring participant safety.
- **Invest in site training:** Comprehensive, role-specific training is provided before trial initiation and reinforcement throughout conduct. Investigators and coordinators understand not just the procedures but the rationale behind requirements.
- **Establish quality culture:** There is an effort to move beyond compliance-focused monitoring toward collaborative quality partnerships with sites. Share Transparent quality metrics are shared and sites demonstrating excellence are recognized.

These four challenge areas (ethical compliance, recruitment and retention, diversity, and data quality) are deeply interconnected. Addressing the challenge areas effectively requires integrated strategies that recognize how improvements in one area can positively affect other areas, while failures cascade across the entire trial enterprise.

# Applications of technology and innovation in clinical trials

The rapid advances in technology have fundamentally transformed clinical trial design and implementation. This section examines four major technological domains (data science, real-world evidence, artificial intelligence, personalized medicine, and digital health) that are reshaping modern clinical trials. **Figure 2** provides a hierarchical overview of these technology applications, illustrating how each domain contributes to improved trial efficiency and outcomes.

## Data science and clinical trial data analysis

As the scale and complexity of clinical trials continue to grow, the application of data science has become crucial. Clinical trials generate vast amounts of diverse data, including patient health conditions, treatment responses, laboratory test results, and medical images. Data science helps researchers better understand these data, uncover potential patterns and trends, which enhance the quality and efficiency of decision-making in clinical trials [92].

Applications of data science in clinical trials include, but are not limited to the following [93–95]:

- **Data Integration and Management:** Modern database technologies, such as big data platforms and cloud computing, are used to integrate data from different sources, while ensuring data quality and consistency.
- **Statistical Analysis:** Advanced statistical models and methods are used to analyze clinical trial data, evaluating drug efficacy and safety. Techniques, such as multivariable analysis and regression analysis, are used to extract meaningful conclusions from complex data sets.

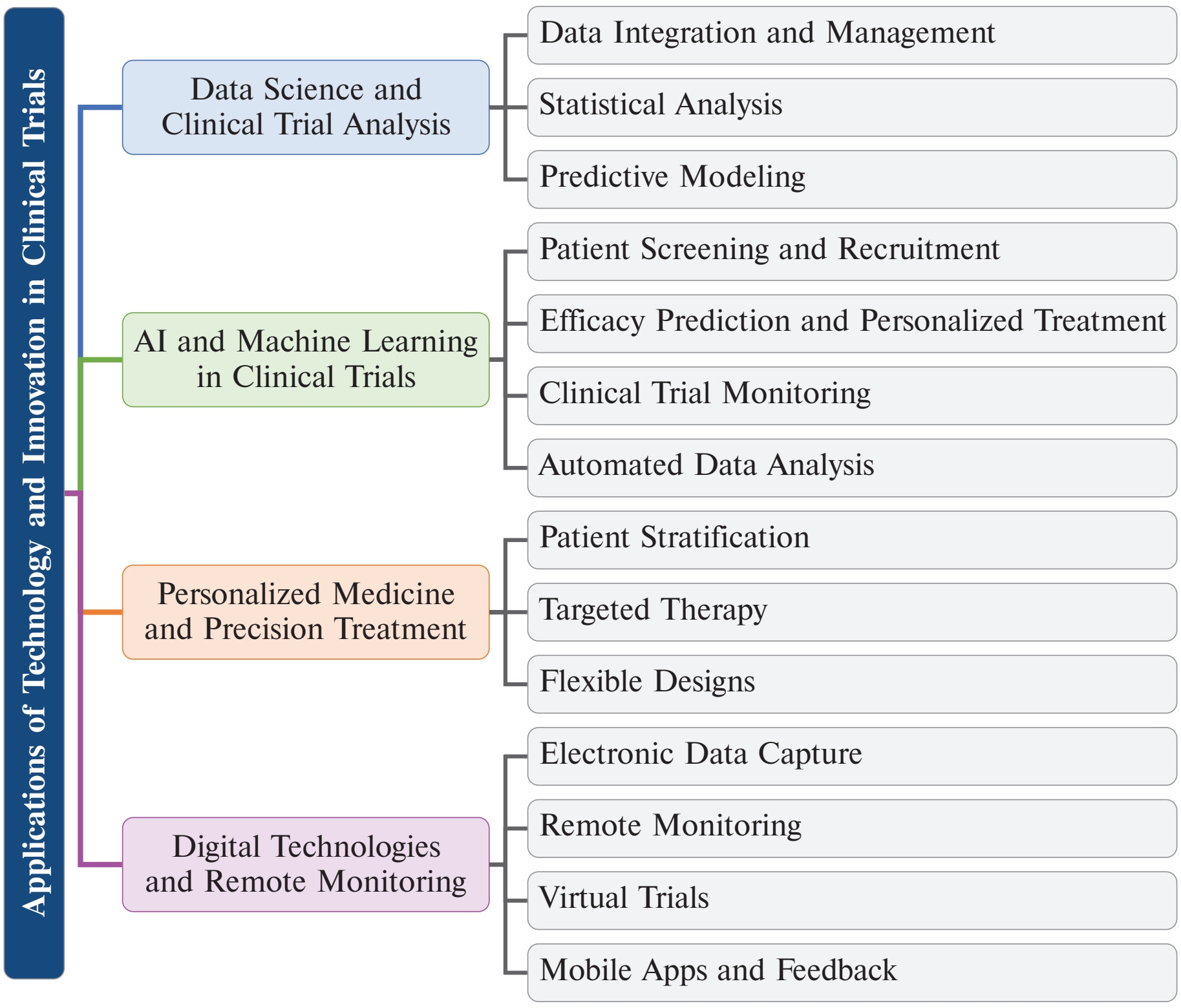

**Figure 2** Hierarchical overview of applications of technology in clinical trials.

- **Predictive Modeling:** Machine learning and data mining techniques are used to build predictive models to assess patient treatment responses and potential side effects, aiding in optimizing decisions early in the trial process.

Data science not only improves the precision of data analysis but also reduces human bias, ensuring that the results of clinical trials are more reliable.

## Real-world evidence and pragmatic trials

Real-world evidence (RWE) derived from real-world data (RWD) is increasingly recognized as a valuable complement to traditional randomized controlled trials (RCTs) [15]. RWD includes data from electronic health records (EHRs), claims databases, patient registries, and wearable devices, providing insights into treatment effectiveness in diverse patient populations under routine clinical practice conditions.

The integration of RWE into drug development offers several advantages:

- **Post-marketing Surveillance Enhancement:** RWE enables continuous monitoring of drug safety and effectiveness after approval, complementing Phase IV studies with larger and more diverse patient populations.
- **External Control Arms:** RWD can provide historical or concurrent control groups when randomization is challenging or unethical, reducing trial costs and accelerating development timelines.
- **Label Expansion Studies:** RWE supports the investigation of new indications or patient subgroups, informing regulatory decisions for label expansions.
- **Pragmatic Trial Design:** Pragmatic trials conducted in real-world settings bridge the gap between efficacy (as demonstrated in controlled trials) and effectiveness (as observed in clinical practice).

Regulatory agencies, including the US FDA and EMA have issued guidance on the use of RWE in regulatory decision-making [96]. The 21st Century Cures Act in the United States explicitly encourages the use of RWE to support drug approvals and post-market requirements. However, challenges remain in ensuring data quality, standardization, and addressing confounding factors inherent in observational data.

## Artificial intelligence and machine learning in clinical trials

Artificial intelligence (AI) and machine learning (ML) technologies are accelerating applications in clinical trials. By leveraging large-scale data sets and powerful computing capabilities, AI and ML assist researchers in analyzing clinical data more accurately, optimizing trial designs, and improving decision-making efficiency. AI technologies have a role not only in the early stages of drug discovery but also throughout the entire clinical trial process [97–99].

Key applications of AI and ML in clinical trials include the following [100–102]:

- **Patient Screening and Recruitment:** AI can analyze historical data, electronic medical records (EMRs), and genomic data to identify potential patients that meet trial criteria, optimizing the patient recruitment process.

- **Efficacy Prediction and Personalized Treatment:** AI models can predict drug efficacy and side effects based on the clinical characteristics and genetic information of patients, providing personalized treatment plans for each patient.
- **Clinical Trial Monitoring:** AI can monitor the progress of clinical trials in real-time, analyze patient responses, detect potential adverse events, and provide early warnings, ensuring trial safety.
- **Automated Data Analysis and Reporting:** Machine learning algorithms can automate the processing of large and complex data, enhancing efficiency and accuracy in data analysis, while minimizing human error [103].

As AI and ML technologies continue to develop, these technologies are expected to significantly improve the efficiency, reduce costs, and accelerate the drug development process in clinical trials.

Practical Applications and Case Examples: Several AI applications have demonstrated tangible impact in clinical trials [12]. Companies, such as Recursion Pharmaceuticals (Salt Lake City, UT, USA), have utilized deep learning to identify drug candidates for rare diseases, significantly accelerating the lead optimization phase. AI-powered platforms have reduced enrollment timelines in patient recruitment by 30–50% through automated eligibility screening of EHRs. The COVID-19 pandemic accelerated adoption of AI-powered trial monitoring with platforms enabling real-time safety signal detection across decentralized trials.

Limitations and Challenges: Despite these advances, significant barriers remain in AI adoption for clinical trials:

- **Data Privacy and Security:** AI systems require access to sensitive patient data, raising concerns about HIPAA/GDPR compliance and data breaches. Federated learning approaches are emerging but remain technically challenging to implement at scale.
- **Regulatory Uncertainty:** FDA and EMA guidelines on AI/ML in clinical trials are still evolving, creating uncertainty for sponsors regarding validation requirements and approval pathways.
- **Algorithm Bias and Transparency:** Machine learning models trained on historically biased datasets may perpetuate health disparities. The "black box" nature of deep learning models poses challenges for regulatory review and clinical interpretation.
- **Standardization and Interoperability:** Lack of standardized data formats across healthcare systems limits the scalability of AI solutions in multicenter trials.

## Personalized medicine and precision treatment in clinical trial design

Personalized medicine and precision treatment are major breakthroughs in the current medical research field. By analyzing the genome, environment, and lifestyle of a patient, personalized medicine aims to tailor the most effective treatment for each individual patient [104]. Clinical trial designs are also evolving to align with personalized medicine, ensuring that each patient receives the most suitable treatment.

In the context of personalized medicine, clinical trial design differs significantly from traditional trials. Key aspects of personalized clinical trial design include the following[105–108]:

- **Patient Stratification:** Patients are categorized into different subgroups based on genetic features and pathologic characteristics, and each subgroup is analyzed and assessed independently. This stratification improves the accuracy of trial results and ensures that different patient types benefit from the appropriate treatments.
- **Targeted Therapy:** Clinical trials are designed around targeted drugs aimed at specific genes or biomarkers. For example, some cancer patients may respond well to targeted therapies, while other cancer patients may not respond. Targeted therapy trials focus more on individual differences.
- **Flexibility in Clinical Trial Design:** Personalized clinical trials require more flexible designs, allowing for adjustments based on early assessments of treatment efficacy and mid-trial modifications to therapy protocols.

Personalized medicine clinical trial designs will make treatments more targeted, reduce unnecessary side effects, and improve treatment outcomes.

## Digital technologies and remote monitoring in clinical trials

Digital technologies and remote monitoring are increasingly applied in clinical trials, particularly during the COVID-19 pandemic, during which the use of telemedicine and remote monitoring significantly increased [109]. Digital technologies not only enhance the efficiency and quality of clinical trials but also ensure the smooth progression during exceptional circumstances like a pandemic.

Applications of digital technologies in clinical trials include the following [97, 110, 111]:

- **Electronic Data Capture (EDC)**: Electronic data collection and storage systems reduce errors and delays associated with traditional paper records, which accelerates data processing and improves accuracy.
- **Remote Monitoring and Patient Tracking:** Wearable devices and mobile health apps enable real-time monitoring of patient health, allowing researchers to adjust treatment plans promptly. Remote monitoring technologies also reduce patient participation costs, making trials more flexible and convenient.
- **Virtual Clinical Trials:** Some clinical trials are conducted entirely online or remotely with the ongoing development of virtual technologies, eliminating the need for patients to be physically present at trial sites, thereby reducing costs and increasing patient enrollment.
- **Mobile Applications and Patient Feedback:** Patients can submit symptoms, treatment feedback, and other health data at any time through mobile applications, which researchers can analyze in real-time, enabling timely reactions and adjustments to the trial protocol.

The combination of digital technologies and remote monitoring not only provides greater convenience for patients but also ensures that clinical trials can be conducted more efficiently and safely, enhancing the overall quality and feasibility of the trial process.

**Figure 2** summarizes the technology domains discussed in this section, illustrating how data science, AI, personalized medicine, and digital technologies collectively transform clinical trial practice. These innovations address many of the challenges identified in the 'Challenges and response strategies in clinical trials' section particularly in improving recruitment efficiency, enhancing data quality, and enabling more diverse patient participation.

## Future perspectives and challenges

The challenges identified in the 'Challenges and response strategies in clinical trials' section (ethical compliance, participant recruitment, diversity, and quality control) continue to shape the evolution of clinical trials. This section explores how emerging technologies and therapeutic approaches are being developed specifically to address these persistent obstacles, while also introducing new considerations for trial design and regulatory frameworks.

As medical science and technology continue to advance, clinical trials are evolving to address these challenges and embrace emerging opportunities. **Figure 2** illustrates the hierarchical overview of technology applications discussed.in the 'Applications of technology and innovation in clinical trials' section which provide the foundation for addressing many of the challenges outlined earlier. The following subsections examine future directions in clinical trial design, the impact of novel therapeutic approaches, and the importance of regulatory adaptation and international collaboration.

## Future directions in clinical trials

With the continuous advances in technology and medicine, the design and execution of clinical trials are also evolving. Future clinical trials will increasingly focus on precision, personalization, and flexibility to meet the changing medical needs and treatment approaches.

There are several important future directions in clinical trials:

- **Precision Medicine-driven Clinical Trials** [112, 113]: With the rapid development of genomics, proteomics, metabolomics, transcriptomics, and epigenomics future clinical trial designs will emphasize individualized and precise treatments. By analyzing the genetic information, environmental factors, and lifestyle habits of patients, clinical trials will be able to tailor drug treatment regimens for different patient groups to improve efficacy and reduce side effects.
- **The Rise of Virtual and Remote Clinical Trials** [114–117]: Due to technological advances, especially the rapid development of information and communication technologies, future clinical trials may rely more on virtual platforms and remote monitoring technologies. For example, patients can provide treatment feedback and health data through smart devices and mobile applications, allowing researchers to analyze the data in real time and make adjustments, thus increasing the flexibility and patient participation in clinical trials.
- **Data-Driven Trial Optimization** [118, 119]: With the application of big data, AI, and machine learning, future clinical trials will place more emphasis on the rapid collection, analysis, and application of data. AI and big data can help optimize trial designs, predict patient responses, identify potential side effects, and accelerate trial progress through intelligent decision-making tools.

These developments will improve the efficiency and accuracy of clinical trials, shorten the time to market for drugs, and enhance patient treatment outcomes.

## Impact of Emerging Therapies on Clinical Trial Design

Emerging therapies, such as gene therapy and immunotherapy, are redefining modern medical treatment approaches. These therapies typically involve highly personalized treatments, which poses new challenges and demands on traditional clinical trial designs.

- **Challenges in Gene Therapy** [120–122]: Gene therapy involves altering or replacing the genes of a patient to treat diseases. These therapies typically target a small group of specific patients, so clinical trials need to screen patients based on specific genetic profiles or pathologic characteristics. Clinical trial designs for gene therapy are often more complex and must consider the effectiveness of gene transfer technologies, long-term safety, and potential immune reactions.

For example, Luxturna (voretigene neparvovec), approved by the US FDA in December 2017 for inherited retinal dystrophy caused by biallelic RPE65 mutations, exemplifies the unique challenges of gene therapy trials [123]. The Phase III trial enrolled only 31 participants due to the rare nature of the disease, required novel functional vision endpoints rather than traditional visual acuity measures, and necessitated long-term follow-up to assess durability of the one-time treatment. Similarly, Zolgensma (onasemnogene abeparvovec), which is indicated for spinal muscular atrophy, demonstrated the importance of early intervention with the pivotal trial showing that single-dose gene replacement therapy could achieve remarkable motor milestone improvements in infants who would otherwise face progressive motor neuron loss [124]. These cases highlight how gene therapy trials must accommodate small patient populations, develop disease-specific endpoints, and implement extended monitoring protocols.

- **Immunotherapy in Clinical Trial Design** [125–127]: Immunotherapy involves activating or suppressing the immune system to treat cancer or other immune-related diseases. A key characteristic of immunotherapy is the large individual variability in efficacy, so clinical trials often need to be multicenter, multinational, and long-term to better assess efficacy and understand the diversity of immune responses. In addition, immunotherapy clinical trials face challenges in managing side effects, such as immune-related adverse events, making patient monitoring and side effect evaluation crucial in the trial design.

The development of pembrolizumab (Keytruda) pioneered the basket trial approach, in which patients are enrolled based on molecular markers rather than tumor type. In June 2020, the US FDA approved pembrolizumab for tumor mutational burden-high (TMB-H) solid tumors based on data from KEYNOTE-158, which enrolled patients across nine different tumor types; this was the first approval based on this biomarker [128]. CAR-T therapies, such as Kymriah and Yescarta, present unprecedented challenges, including autologous cell manufacturing logistics requiring approximately 3 weeks, management of cytokine release syndrome, and the need for specialized treatment centers with intensive care capabilities [17]. These immunotherapy trials have fundamentally changed clinical trial design by demonstrating that biomarker-driven, tumor-agnostic approaches can achieve regulatory approval.

Clinical trials for emerging therapies not only require innovative designs but also more precise patient selection and treatment monitoring mechanisms to ensure the safety and efficacy of these new therapies. The experiences with Luxturna, Zolgensma, pembrolizumab, and CAR-T therapies provide valuable lessons for future trial design in this rapidly evolving field.

## Regulatory changes and global collaboration in clinical trials

International collaboration in clinical trials has become an important trend with the deepening of globalization and technological advancements. However, regulatory differences and changing frameworks across countries pose distinct challenges for global clinical trials.

- **Challenges in Regulatory Changes** [129–132]: Clinical trial regulations and policies evolve with the emergence of new technologies and therapies. For example, as genomics and precision medicine advance, the regulatory agencies in many countries are updating the policies to accommodate these emerging technologies. In addition, regulatory bodies are also revising regulations on data privacy, patient protection, and drug approval processes to keep up with global collaboration.

**Expedited Approval Pathways:** Regulatory agencies have developed accelerated pathways to address unmet medical needs [19, 20]. The US FDA Breakthrough Therapy Designation, granted to > 300 drug development programs since 2012, enables intensive guidance and rolling review for promising therapies. Accelerated approval allows marketing based on surrogate endpoints with confirmatory trials required post-approval, a pathway used for approximately 25% of novel drug approvals in oncology. The EMA Priority Medicines (PRIME) scheme provides similar benefits for European submissions.
**Gene and Cell Therapy Regulations:** The regulatory framework for advanced therapies has evolved rapidly. The US FDA 2020 guidance on gene therapy manufacturing established quality standards for viral vector production. The Regenerative Medicine Advanced Therapy (RMAT) designation provides expedited development for regenerative medicine products. However, challenges persist in harmonizing long-term follow-up requirements; the US FDA recommends 15-year monitoring for gene therapies, while the EMA guidance varies by product class.
**Decentralized Trial Regulations:** The COVID-19 pandemic prompted regulatory flexibility for decentralized trials (DCTs). The US FDA 2023 guidance on DCTs clarified requirements for remote consent, direct-to-patient drug shipment, and telemedicine visits. However, international harmonization remains limited with varying acceptance of DCT elements across regulatory jurisdictions.

- **Global Collaboration in Clinical Trials** [133–135]: Globalization enables multinational companies and research institutions to conduct clinical trials simultaneously in multiple countries. Multinational clinical trials can accelerate drug approval processes, expand sample sizes, and improve the representativeness of results. However, cross-border collaborations face challenges in terms of cultural, legal, and ethical issues. Different countries have varying requirements for trials. Standards for data management and ethical review differ, presenting higher demands for global cooperation. Strengthening collaboration between countries, harmonizing regulatory standards, and simplifying approval procedures will be key to the development of global clinical trials.

Global clinical trials not only accelerate drug development and approval but also facilitate the sharing of medical technologies between different regions, advancing global public health efforts.

# Conclusion

## Summary of key findings

This review has examined the clinical trial landscape across the following four dimensions: trial methodology; technological innovation; regulatory frameworks; and emerging therapeutic challenges. Our analysis reveals that while traditional phase-based trial structures remain foundational, adaptive designs, real-world evidence integration, and AI-enabled processes are fundamentally reshaping trial conduct. The case studies of gene therapies (Luxturna and Zolgensma) and immunotherapies (pembrolizumab and CAR-T) demonstrate that novel therapeutic modalities require correspondingly innovative trial designs.

## Recommendations for stakeholders

Based on our comprehensive review, we offer the following actionable recommendations:

**For Trial Designers and Sponsors:**

- Implement adaptive trial designs with pre-specified interim analyses to improve efficiency and reduce resource waste from failed trials

- Integrate real-world data sources for external control arms in rare disease trials in which randomization is impractical
- Develop biomarker-driven enrollment strategies to enable precision medicine approaches and basket/umbrella trial designs
- Establish decentralized trial capabilities including remote consent, direct-to-patient drug delivery, and wearable device integration for continuous monitoring

**For Regulatory Agencies:**

- Harmonize international guidelines for decentralized trials to facilitate global clinical development programs
- Develop clear validation frameworks for AI/ML algorithms used in trial design, patient selection, and safety monitoring
- Establish standardized requirements for real-world evidence quality to enable consistent regulatory decision-making
- Create expedited pathways specifically designed for platform trials testing multiple interventions

**For Policymakers and Healthcare Systems:**

- Invest in clinical trial infrastructure, including EHR, interoperability to facilitate patient identification and data collection
- Address recruitment barriers through community engagement programs targeting underrepresented populations
- Support workforce development in clinical trial methodology, data science, and regulatory science

## Future outlook

The convergence of technological innovation, regulatory evolution, and therapeutic advancement will continue to transform clinical trials. Success in this evolving landscape will require stakeholders to embrace adaptive methodologies, while maintaining rigorous scientific standards. The ultimate goal is to accelerate the development of safe and effective therapies to improve patient outcomes worldwide, and remains unchanged.

## Data availability statement

This is a review article. No new data were generated or analyzed in this study. All data discussed are from previously published studies cited in the reference list.

## Author contributions

Conceptualization: T.W., Z.J.; Literature search and analysis: T.W., M.L., B.P., X.S., C.Z., X.S., Q.N., J.L., S.C., K.C., M.L., P.F., Z.B., Y.W., Y.Z., C.F., L.K.Q.Y., Z.Q., R.B.; Writing—original draft: T.W., M.L., B.P.; Writing—review and editing: Z.J., T.W.; Supervision: Z.J. All authors have read and agreed to the published version of the manuscript.

## Funding

This research received no external funding.

## Acknowledgments

Google Gemini and ChatGPT were used to assist in the preparation of Graphical abstract in this manuscript. All generated content was reviewed, edited, and approved by the authors, who take full responsibility for the accuracy and integrity of the figures.

## Conflict of interest

The authors declare that there are no conflicts of interest.